# Ion relaxation dynamics in $0.5Li_2O$-$0.5Na_2O$-$2B_2O_3$ ($LiNaB_4O_7$) glasses

Rahul Vaish and K.B.R. Varma*

Materials Research Centre, Indian Institute of Science, Bangalore, India.

*Corresponding Author; E-Mail: kbrvarma@mrc.iisc.ernet.in;

FAX: 91-80-23600683; Tel. No: 91-80-22932914

**Abstract**

The frequency and temperature dependence of the dielectric constant, electric modulus and electrical conductivity of the transparent glasses in the composition $0.5Li_2O$-$0.5Na_2O$-$2B_2O_3$ (LNBO) were investigated in the 100 Hz - 10 MHz frequency range. The dielectric data have been analyzed using Cole-Cole equation with an addition of the conductivity term. The electrical conductivity was found to be obeying Jonscher's universal law. The dielectric constant and the loss for the as-quenched glasses increased with increasing temperature, exhibiting anomalies in the vicinity of the glass transition and crystallization temperatures. The bulk DC conductivity at various temperatures was extracted from the electrical relaxation data. Conductivity plots were not found to follow Summerfield scaling due to mixed alkali effect. The imaginary part of electric modulus spectra was modeled using an approximate solution of Kohlrausch-Williams-Watts (KWW) relation. The stretching exponent, *ß*, was found to be 0.56 and temperature independent.

## 1.Introduction

Ferroelectric glass-ceramics have been studied extensively owing to their fascinating optical and electronic properties [1-4]. The transparent glasses embedded with polar nano/micro crystals could be tailored to exhibit non-linear optical, piezoelectric and pyroelectric properties depending on the crystallite size, volume fraction and microstructure. These are also of technological prominence because of the fact that their microstructure could be engineered precisely for specific applications via controlled heat-treatment of the as-quenched glasses. Glass-ceramics are easy to fabricate and are often cost effective as compared to their single crystalline counterparts.

Borate-based single crystals and their glass counterparts are one of the important class of materials for photonic applications. These materials have become attractive because of their superior characteristics such as: (I) large non-linear optical coefficients, (II) high laser induced damage thresholds, (III) wide transparency range for operating wavelengths, (IV) easy to fabricate, (V) moderate melting points, (VI) good chemical and mechanical stability and (VII) low material cost [5]. Various borate-based single crystals, including $LiB_3O_5$ [6], $CsLiB_6O_{10}$ [7], $SrB_4O_7$ [8] have been investigated and reported to be promising from their physical properties view point. These compounds are very attractive particularly for their applications in non-linear photonics. The aforementioned noncentrosymmetric borate single crystals are the most promising among the other noncentrosymmetric compounds for non-linear optical based devices. Unfortunately growing large single crystals of these materials from their melts is difficult because of their high viscosities. Therefore, we have been exploring alternate ways of obtaining transparent materials. One of the routes that drew our attention was the glass-ceramic.

Amongst various borate-based materials, $Li_2B_4O_7$ single crystals and glass ceramics have been reported to be promising from their above mentioned functional properties view point [9]. It has good piezoelectric properties, wide transmission window and large optical damage threshold. However, $Li_2B_4O_7$ has high ionic conductivity which causes hindrance to the piezoelectric and pyroelectric properties. Attempts were made to replace partially $Li^+$ by $Na^+$ and visualize its physical properties. Structural details of $LiNaB_4O_7$ single crystals have been reported [10]. It was found to crystallize in the orthorhombic system associated with non-centrosymmetric space group. Since it is polar, it deserves much attention from its dielectric and electrical transport properties point of view as these properties have direct influence on its pyroelectric and piezoelectric characteristics. The present investigations aim at fabricating transparent glasses and glass-ceramics of the above compositions and make an attempt to understand their electrical transport and non-linear properties. To begin with, glasses in the composition $0.5Li_2O$- $0.5Na_2O$-$2B_2O_3$ (which on heating at appropriate temperatures yielded crystalline $LiNaB_4O_7$ phase) have been investigated for their dielectric and electrical conductivity properties over the range of temperatures and frequencies that are normally of interest in the applications of these materials. The experimental data have been analyzed using three different formalisms: complex dielectric constant, complex electric modulus and electrical conductivity. Transformations from one formalism to another may throw light on the ion dynamics in the glasses.

## 2. Experimental:

Transparent LNBO glasses were fabricated via the conventional melt-quenching technique. For this, $Na_2CO_3$ (99.95% Aldrich), $Li_2CO_3$ (99.9% Merck) and $H_3BO_3$ (99.9% Merck) were mixed and melted in a platinum crucible at 1050$^oC$ for 30 min. Melts were quenched by pouring on a steel plate and pressed with another plate to obtain 1-1.5 mm thick glass plates. All these samples were annealed at 450$^oC$ (for 6h) which is well below the glass-transition temperature. X-ray powder diffraction study was performed at room temperature on the as-quenched samples to confirm their amorphous nature. The glassy nature of the as-quenched samples was confirmed by subjecting the samples to non-isothermal Differential Scanning Calorimetric (Perkin Elmer, Diamond DSC) studies, in the 350-700$^oC$ temperature range.

The capacitance and dielectric loss ( $D = \varepsilon_r'' / \varepsilon_r'$ ) were monitored as functions of both frequency (100 Hz-10 MHz) and temperature (40$^oC$-650$^oC$), using a HP4194A impedance/gain phase analyzer at a signal strength of 0.5 $V_{rms}$. Based on these data the real and imaginary parts of the dielectric constant were calculated by taking the dimensions and electrode geometry of the sample into account. For this purpose the as-quenched glass samples were painted with silver and silver epoxy was used to bond silver leads to the sample and thus formed the parallel plate capacitor geometry.

## 3. Results and Discussion

The XRD pattern that is obtained for the as-quenched sample ($0.5Li_2O$-$0.5Na_2O$-$2B_2O_3$ in molar ratio) shown in Fig. 1 confirms its amorphous state. The DSC trace, recorded in the 350- 700 $^oC$ temperature range for the as-quenched glass plates is shown in Fig. 2. It exhibits the glass transition (endotherm, 440$^{o}$C) and an exotherm at 544$^{o}$C associated with the crystallization. The variation of the dielectric constant ($\varepsilon_r^{'}$) with frequency (100 Hz – 10 MHz) of measurement for LNBO glass-plates at different temperatures is shown in Fig. 3 (a). At all the temperatures under investigation, $\varepsilon_r^{'}$ decreases with increase in frequency. The decrease is significant, especially at low frequencies, which may be associated with the mobile ion polarization combined with electrode polarization. The low-frequency dispersion of $\varepsilon_r^{'}$ gradually increases with increase in temperature due to an increase in the electrode polarization as well as the thermal activation associated with $Li^{+}$ and $Na^{+}$ ions in the LNBO glasses. The electrode polarization is significant at high temperatures (200-400$^{o}$C) and masks the bulk response of the glasses in the low frequency regime. When the temperature rises, the dielectric dispersion shifts towards higher frequencies. The complex dielectric constant ($\varepsilon_r^{*}$) in LNBO glasses could be rationalized using the Cole-Cole equation [11]:

$$\varepsilon_r^{*} = \varepsilon_\infty + \frac{\varepsilon_s - \varepsilon_\infty}{1+(i\omega\tau)^{1-\alpha}} \tag{1}$$

where $\varepsilon_s$ is the static dielectric constant, $\varepsilon_\infty$ is the high frequency value of the dielectric constant, *?* (=2p*f*) is the angular frequency, *t* is the dielectric relaxation time and *a* is a measure of distribution of relaxation times with values ranging from 0 to 1. For an ideal Debye relaxation, *a* = 0 and for *a* > 0 the relaxation has a distribution of relaxation times. The real part of the dielectric constant ($\varepsilon_r^{'}$) from eq.1 could be expressed as

$$\varepsilon_r^{'} = \varepsilon_\infty + \frac{(\varepsilon_s - \varepsilon_\infty)[1 + (\omega\tau)^{1-\alpha}\sin(\alpha\pi/2)]}{1 + 2(\omega\tau)^{1-\alpha}\sin(\alpha\pi/2) + (\omega\tau)^{2-2\alpha}} \quad (2)$$

The experimental data on the variation of $\varepsilon_r^{'}$ with frequency could not be fitted perfectly using Eq. 2 in the entire frequency range since the Cole-Cole equation predicts nearly constant $\varepsilon_r^{'}$ in the low frequency regime, which is not true in the present case. This is due to the fact that the electrode/space charge polarization is dominant at low frequencies as depicted in Fig. 3 (a). The above observations necessitate the inclusion of the electrical conductivity term in the Cole-Cole equation to rationalize the $\varepsilon_r^{'}$ versus frequency behavior of LNBO glasses in the whole frequency range. After adding the term that also reflects the electrode/space charge polarization in the Eq. 2, one arrives at [12]

$$\varepsilon_r^{'} = \varepsilon_\infty + \frac{(\varepsilon_s - \varepsilon_\infty)[1 + (\omega\tau)^{1-\alpha}\sin(\alpha\pi/2)]}{1 + 2(\omega\tau)^{1-\alpha}\sin(\alpha\pi/2) + (\omega\tau)^{2-2\alpha}} + \frac{\sigma_2}{\varepsilon_o\omega^s} \quad (3)$$

where $s$ (0, 1) is a constant. It signifies the distribution of the carrier polarization. For an ideal complex conductivity, the value of $s$ is 1**.** $s_2$ is the conductivity, a contribution from the space charges. Solid lines in Fig. 3 (a) are the fitted curves of the experimental results (100 Hz-10 MHz) according to Eq. 3. The parameters that are obtained from the best fit at various temperatures are presented in Table I. It is to be noted that the value of $s$ increases with increasing temperature (Table I), indicating that the space charge related polarization has significant dispersion at higher temperature. In order to further elucidate the dielectric relaxation in LNBO glasses, it is important to estimate the activation energy associated with the relaxation process. The activation energy involved in the relaxation process of ions could be obtained from the temperature dependent relaxation time (Table I)

$$\tau = \tau_o \exp\left(E/_{kT}\right) \qquad (4)$$

where $E$ is the activation energy associated with the relaxation process, $\tau_o$ is the pre-exponential factor, $k$ is the Boltzmann constant, and $T$ is the absolute temperature. Fig. 3 (b) depicts the plot of ln (t) versus 1000/$T$ along with linear fit (solid line) to the above equation (Eq. 4). The value that is obtained for $E$ is 1.32eV, which is ascribed to the motion of $Li^+$ and $Na^+$ ions in the glass matrix. This relatively large value of activation energy is due to the mixed alkali effect and in good agreement with that reported in the literature [13]. The variation of the dielectric loss ($D$) with the frequency at various temperatures is shown in Fig. 4. The loss decreases with increase in frequency at different temperatures ($200^oC$-$400^oC$). However, it increases with increase in temperature, which is attributed to the increase in electrical conductivity of the glasses. The relaxation peak is not encountered at any temperature under study because of dominant DC conduction losses due to high cation mobility.

To get further insight into the dielectric behavior, the temperature variation of the real part of the dielectric constant of the glass at various frequencies is studied (Fig. 5 (a)). The dielectric constant is found to increase gradually in the 30-150 $^oC$ temperature range and subsequently increases rapidly up to about 400 $^oC$ at all the frequencies under study (inset of Fig. 5(a)). This dielectric anomaly is associated with the glass-transition temperature of LNBO glasses. It is well known that the physical properties (heat capacity, viscosity, and thermal expansion coefficient) of a glass often changes abruptly while passing through its glass-transition. When the viscosity of the glass abruptly decreases in the glass-transition region, the reaction elements such as dipoles and ions easily respond to the external electric field and the dielectric constant increases. On further heating, the dielectric constant increases rapidly and exhibits a peak around 480$^{o}$C. This temperature range is not consistent with that of $T_g$ and $T_{cr}$ for the as-quenched LNBO glasses (Fig. 2) because of different time scales involved. A sharp peak around the crystallization temperature (480$^{o}$C) of LNBO glasses is attributed to an increase in the interfacial polarization. During the crystallization process the interfaces (that would get created) between the glassy regions and the crystallites having different dielectric constants and conductivities could be the origin of charge accumulation and hence interfacial polarization. The sharp rise in the dielectric constant might also be due to the ions that indulge in rapid movement to transform from a random (glassy) to the ordered (crystalline) state associated with the conduction related polarization during the crystallization. The decrease in dielectric constant above 500 $^{o}$C is due to the reduction in the interfacial polarization and slow ionic movement as the glass is fully crystallized at this temperature. The variation in the dielectric loss ( $D = \varepsilon_r'' / \varepsilon_r'$ ) with the temperature at various frequencies [Fig. 5 (b)] is consistent with that of the dielectric behavior. It is observed that the dielectric loss increases with increase in temperature which is attributed to the increase in conductivity of the glasses.

In order to study the intrinsic electrical relaxation process and to suppress the electrode polarization at low frequency, electric modulus formalism was invoked to rationalize the dielectric response of the present glasses. The complex electric modulus ($M^*$) is defined in terms of the complex dielectric constant ($e$*) and is represented as [14]:

$$M^* = (e^*)^{-1} \qquad (5)$$

$$M^* = M' + iM'' = \frac{\varepsilon_r'}{(\varepsilon_r')^2 + (\varepsilon_r'')^2} + i\frac{\varepsilon_r''}{(\varepsilon_r')^2 + (\varepsilon_r'')^2} \qquad (6)$$

where $M^{'}$ , $M^{''}$ and , $\varepsilon_r^{'}$ , $\varepsilon_r^{''}$ are the real and imaginary parts of the electric modulus and dielectric constants, respectively. The real and imaginary parts of the modulus at different temperatures are calculated using Eq. 6 for the LNBO glasses and depicted in Figs. 6 (a & b), respectively. One would notice from Fig. 6 (a) that at low frequencies, $M^{'}$ approaches zero at all the temperatures under study suggesting the suppression of the electrode polarization. $M^{'}$ reaches a maximum value corresponding to $M_\infty$ = $(\varepsilon_\infty)^{-1}$ due to the relaxation process. It is also observed that the value of $M_\infty$ decreases with the increase in temperature. The imaginary part of the electric modulus (Fig. 6 (b)) is indicative of the energy loss under electric field. The $M^{''}$ peak shifts to higher frequencies with increasing temperature. This evidently suggests the involvement of temperature dependent relaxation processes in the present glasses. The frequency region below the $M^{''}$ peak indicates the range in which cations drift to long distances. In the frequency range which is above the peak, the ions are spatially confined to potential wells and free to move within the wells. The frequency range where the peak occurs is suggestive of the transition from long-range to short-range mobility. The electric modulus ($M^*$) could be expressed as the Fourier transform of a relaxation function ϕ(t):

$$M^* = M_\infty \left[ 1 - \int_0^\infty \exp(-\omega t) \left( -\frac{d\phi}{dt} \right) dt \right] \qquad (7)$$

where the function ϕ(t) is the time evolution of the electric field within the materials and is usually taken as the Kohlrausch-Williams-Watts (KWW) function [15, 16]:

$$\phi(t) = \exp\left[ -\left( t/\tau_m \right)^\beta \right] \qquad (8)$$

where $t_m$ is the conductivity relaxation time and the exponent $\beta$ (0 1] indicates the deviation from Debye-type relaxation. The value of $\beta$ could be determined by fitting the experimental data in the above equations. But it is desirable to reduce the number of adjustable parameters while fitting the experimental data. Keeping this point in view, the electric modulus behavior of the present glass system is rationalized by invoking modified KWW function suggested by Bergman. The imaginary part of the electric modulus ($M^{"}$) is defined as [17]:

$$M"=\frac{M^{"}_{\max}}{(1-\beta)+\frac{\beta}{1+\beta}\left[\beta(\omega_{\max}/\omega)+(\omega/\omega_{\max})^{\beta}\right]} \tag{9}$$

where $M^{"}_{\max}$ is the peak value of the $M^{"}$ and ? $_{max}$ is the corresponding frequency. The above equation (Eq. 9) could effectively be described for $\beta \geq 0.4$. Theoretical fit of Eq. 9 to the experimental data is shown in Fig. 6 (b) as the solid lines. It is seen that the experimental data are well fitted to this model except in the high frequency regime. From the fitting of $M^{"}$ versus frequency plots, the value of $\beta$ was determined and found to be temperature independent. The value of $\beta$ is found to be 0.58 ±0.01 in the 200-350$^{o}$C temperature range. The relaxation time associated with the process was determined from the plot of $M^{"}$ versus frequency. The activation energy involved in the relaxation process of ions could be obtained from the temperature dependent relaxation frequency as:

$$f_{\max} = f_o \exp\left(-\frac{E_R}{kT}\right) \tag{10}$$

where $E_R$ is the activation energy associated with the relaxation process, $f_o$ is the pre-exponential factor, $k$ is the Boltzmann constant and $T$ is the absolute temperature. Fig. 7 shows a plot between ln ($f_{max}$) and 1000/$T$ along with the theoretical fit (solid line) to the above equation (Eq. 10). The value that is obtained for $E_R$ is 1.14 eV, which may be ascribed to the motion of $Li^+$ and $Na^+$ ions and is consistent with the one reported in the literature [18].

In order to reveal the electrical transport mechanism in LNBO glasses, DC conductivity at different temperatures ($\sigma_{DC}(T)$), was calculated from the electric modulus data. The DC conductivity could be obtained using the expression [19]:

$$\sigma_{DC}(T) = \frac{\varepsilon_o}{M_\infty(T) * \tau_m(T)} \left[ \frac{\beta(T)}{\Gamma\left( 1/\beta(T) \right)} \right] \qquad (11)$$

where $\varepsilon_o$ is the free space dielectric constant, $M_\infty$ ($T$) is the reciprocal of high frequency dielectric constant and $\tau_m$ ($T$) (=1/2p$f_{max}$) is the temperature dependent relaxation time. Fig. 7 shows the DC conductivity data obtained from the above expression (Eq. 11) at various temperatures. The relaxation frequency ($f_{max}$) at various temperatures is also depicted in Fig. 7. The activation energy for the DC conductivity is calculated from the plot of ln ($\sigma_{DC}$) versus 1000/$T$ for LNBO glasses. The plot is found to be linear and fitted using the following Arrhenius equation,

$$\sigma_{DC}(T) = B \exp\left( -E_{DC}/kT \right) \qquad (12)$$

where $B$ is the pre-exponential factor, $E_{DC}$ is the activation energy for the DC conduction. The activation energy calculated from the slope of the fitted line is found to be 1.2 eV. This value of activation energy is in good agreement with that obtained for electrical relaxation (1.14 eV) suggesting that both the processes have similar mechanisms. Fig. 8 shows the normalized plots of electric modulus $M^{''}$ versus frequency (*f*) wherein the frequency is scaled by the peak frequency. A perfect overlapping of all the curves on a single master curve is found. This time-temperature superimposition shows that the relaxation dynamics remain unchanged in the temperature range understudy.

AC conductivity at different frequencies and temperatures, was determined by using the dielectric data using the following formula:

$$\sigma_{AC} = \omega\varepsilon_o D\varepsilon_r^{'} \tag{13}$$

where $\sigma_{AC}$ is the AC conductivity at a frequency ? (=2p*f*). The frequency dependence of the AC conductivity at different temperatures is shown in Fig. 9. At low frequency, the conductivity shows a flat response which corresponds to the DC part of the conductivity. At higher frequencies, the conductivity shows a dispersion. It is clear from the figure that the flat region increases with the increase in temperature. In order to analyze the mechanism(s) of conduction in the LNBO glasses, conductivity data are analyzed using Jonscher's law [20]

$$\sigma_{AC} = \sigma_{DC} + A\omega^n \tag{14}$$

where $\sigma_{DC}$ is the DC conductivity, *A* is the temperature dependent constant and *n* is the power law exponent which generally varies between 0 and 1. The exponent *n* represents the degree of interaction between the mobile ions. The present glasses are found to obey the above mentioned universal power law at all the temperatures and frequencies under study. The theoretically fitted lines of Eq. 14 to the experimental data are shown in Fig. 9 (solid lines).

The variation of exponent *n* as a function of temperature is depicted in Fig. 10. It is known that the conductivity mechanism in any material could be understood from the temperature dependent behavior of *n*. To ascertain the electrical conduction mechanism in the materials, various models have been proposed [21]. These models include quantum mechanical tunneling model (QMT), the overlapping large-polaron tunneling model (OLPT) and the correlated barrier hopping model (CBH). According to the QMT model, the value of exponent *n* is found to be 0.8 and increases slightly with increase in the temperature whereas the OLPT model predicts the frequency and temperature dependence of *n*. In the CBH model, the temperature dependent behavior of *n* is proposed. This model states that the charge transport between localized states due to hopping over the potential barriers and predicts a decrease in the value of *n* with the increase in temperature, which is consistent with the behavior of *n* for the glasses under study (Fig.10). This suggests that the conductivity behavior of LNBO glasses can be explained using correlated barrier hopping model.

To explore the possibility of scaling behavior in AC conductivity data, Summerfield scaling was invoked [22]. Summerfield scaling was reported to be valid for wide range of materials from amorphous semiconductors to various ion conducting glasses. This scaling could be achieved by plotting log $[s\ (?)/s_{DC}]$ versus log $[f/(s_{DC.}T)]$. The scaled plots obtained using the above scaling law is shown in Fig. 11. It is clear from the figure that the plots are not perfectly overlapped which indicates invalidity of the above scaling on LNBO glasses. This may be due to mixed alkali effect in these glasses and corroborated with that reported in the literature [23].

Inorder to compare electrical conductivity of the $0.5Li_2O$- $0.5Na_2O$-$2B_2O_3$ glasses to that of $Li_2O$-$2B_2O_3$ (LBO) glasses, the frequency dependence of conductivity plots at $200^oC$ for both the glasses are depicted in Fig. 12. It is clear from the figure that the conductivity is lower for LNBO glasses than that of the LBO glasses in the entire frequency range under study. This is due to the fact that the $Li^+$ and $Na^+$ ions in the LNBO glasses are randomly mixed in all the conduction pathways and each cation creates its own chemical environment [24]. These cation species prefer to migrate via pathways of sites adjusted to its own requirements. The ionic mobility is decreased if the pathways of the respective species interfere with each other. This blocking considerably reduces the conductivity of LNBO glasses in comparison to the corresponding single cation glasses (LBO). These effects result in higher activation energies associated with the conduction in LNBO glasses than those of LBO glasses.

**4. Conclusions**

The ion relaxation dynamics in $0.5Li_2O$-$0.5Na_2O$-$2B_2O_3$ glasses were investigated in the frequency range of 100 Hz-10 MHz. The close agreement between the activation energies associated with DC conductivity and electrical relaxation suggest that the participation of the same ions in both the processes. The behaviour of *n* indicated that the conductivity relaxation could be explained using correlated barrier hopping model. Cole-Cole formalism was invoked to separate bulk dielectric relaxation from electrode effect. The frequency dependent imaginary part of electric modulus indicated non-Debye feature in electrical relaxation. The approximate solution of Kohlrausch-Williams-Watts stretched exponential function was used to rationalize the electric modulus spectra. The value of *ß* was found to be temperature independent. Perfect overlapping of normalized imaginary part of electric modulus plots suggested the relaxation dynamics to be invariant over the investigated temperature range.

**Figure Captions:**

Fig. 1: XRD pattern for the as-quenched pulverized LNBO glasses

Fig. 2: The DSC trace for LNBO glass-plate at a heating rate of 10 K /min

Fig. 3: (a) Frequency dependent dielectric constant plots at various temperatures and solid lines are the fitted curves and (b) ln (t) versus 1000/*T* plot

Fig. 4: Dielectric loss versus frequency plots at various temperatures.

Fig. 5: Temperature dependence of the (a) dielectric constant and (b) loss at various frequencies.

Fig. 6: (a) Real and (b) imaginary parts of the electric modulus as a function of frequency at various temperatures. The solid lines are the theoretical fits.

Fig. 7: Arrhenius plots of DC conductivity and electric relaxation frequency for LNBO glasses

Fig. 8: Normalized spectra of electric modulus vs frequency at various temperatures

Fig. 9: Frequency dependence of AC conductivity at various temperatures and solid lines are theoretical fit

Fig. 10: Exponent (*n*) vs *T* plot for LNBO glasses

Fig. 11: Scaled conductivity spectra of LNBO glasses

Fig. 12: Frequency dependence of conductivity of LBO and LNBO glasses

Table I: Cole-Cole fitting parameters for LNBO glasses

| T (°C) | $e_s$ | $e_8$ | t (μs) | a | $s_2$ (O.m)$^{-1}$ | s | $R^2$ |
|---|---|---|---|---|---|---|---|
| 275 | 177.1 | 9.4 | 0.0004 | 0.415 | 2.73E-8 | 0.826 | 0.9997 |
| 300 | 166.6 | 9.8 | 0.00013 | 0.404 | 1E-7 | 0.879 | 0.9995 |
| 325 | 160.3 | 10.5 | 0.00004 | 0.390 | 4.9E-7 | 0.99 | 0.9998 |
| 350 | 151.5 | 11.1 | 0.00002 | 0.385 | 6.8E-7 | 0.77 | 0.9998 |
| 375 | 143.7 | 12.6 | 5.74E-6 | 0.347 | 2.67E-6 | 0.99 | 0.9999 |
| 400 | 126.3 | 14.1 | 2.07E-6 | 0.292 | 4.72E-6 | 0.99 | 0.9950 |

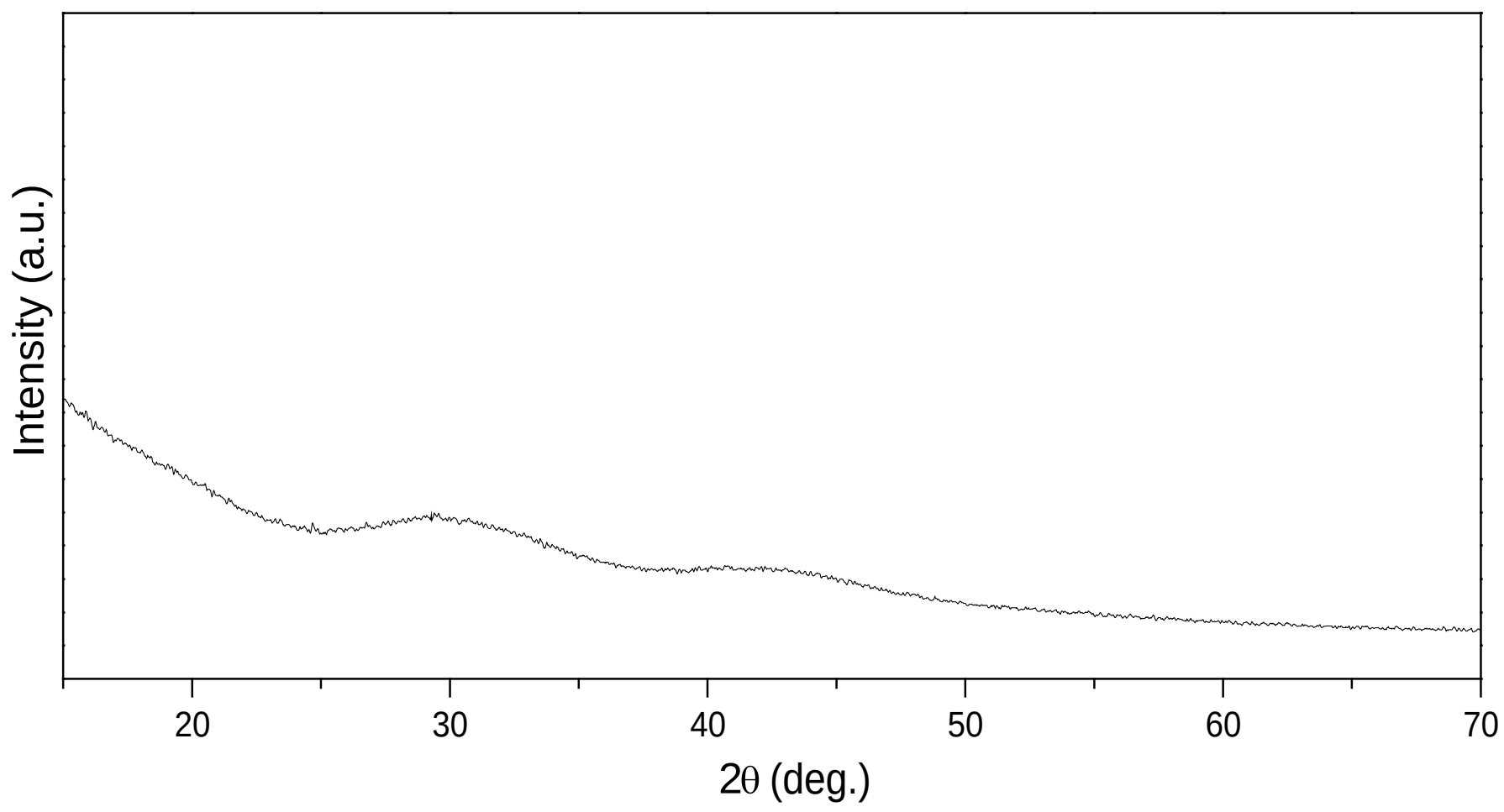

Fig. 1: XRD pattern for the as-quenched pulverized LNBO glasses

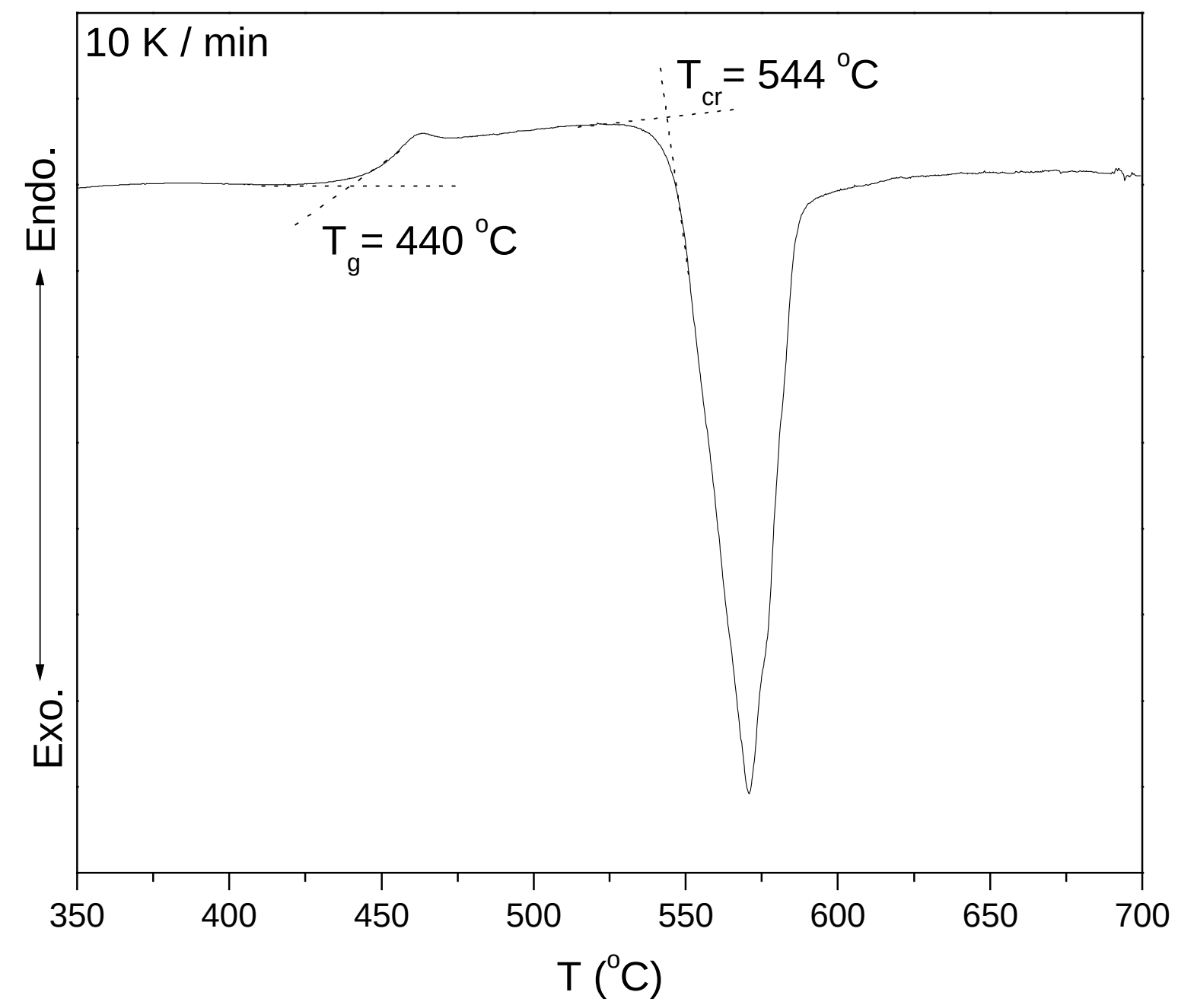

Fig. 2: The DSC trace for LNBO glass-plate at a heating rate of 10 K /min

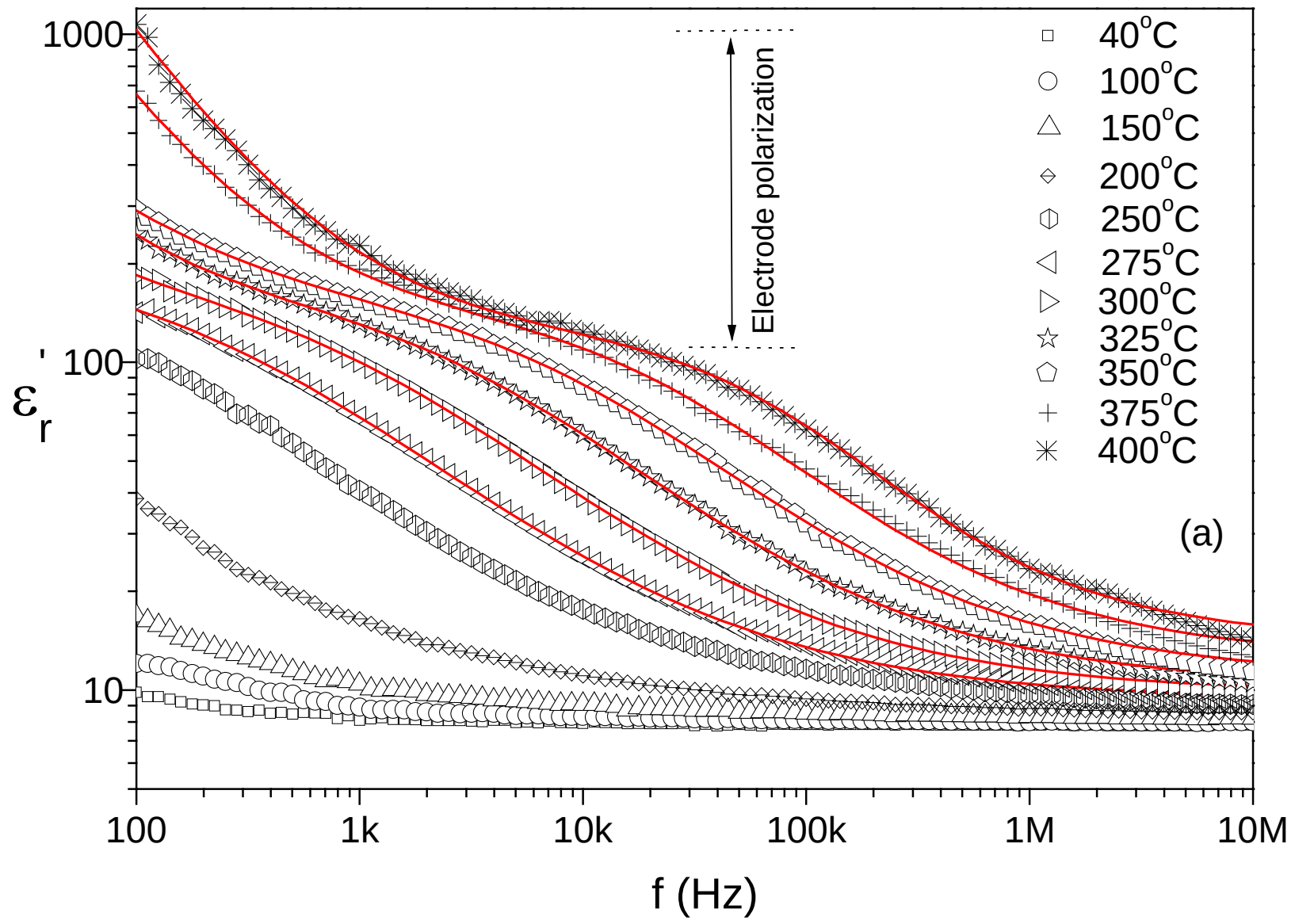

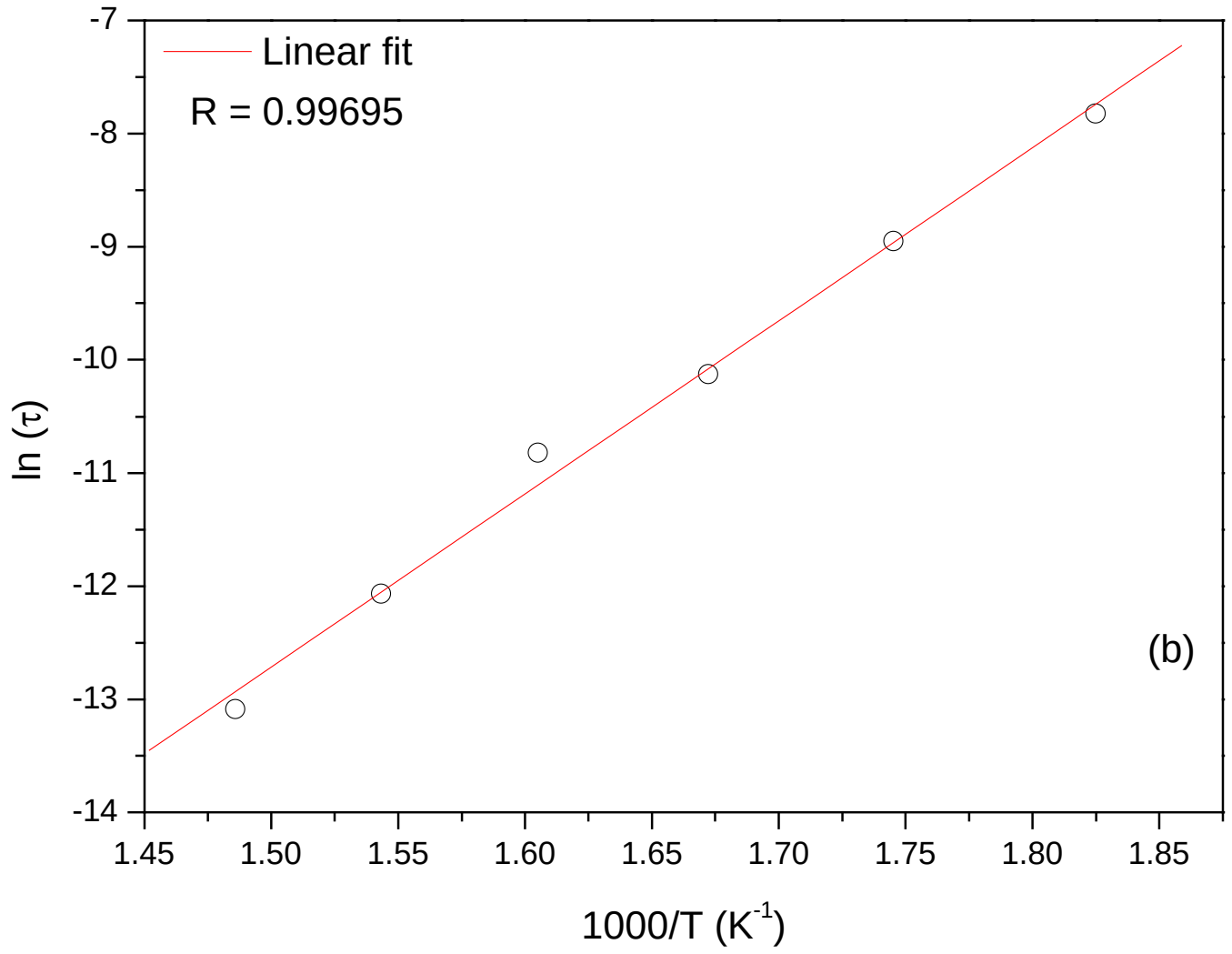

Fig. 3: (a) Frequency dependent dielectric constant plots at various temperatures and solid lines are the fitted curves and (b) ln (t) versus 1000/*T* plot

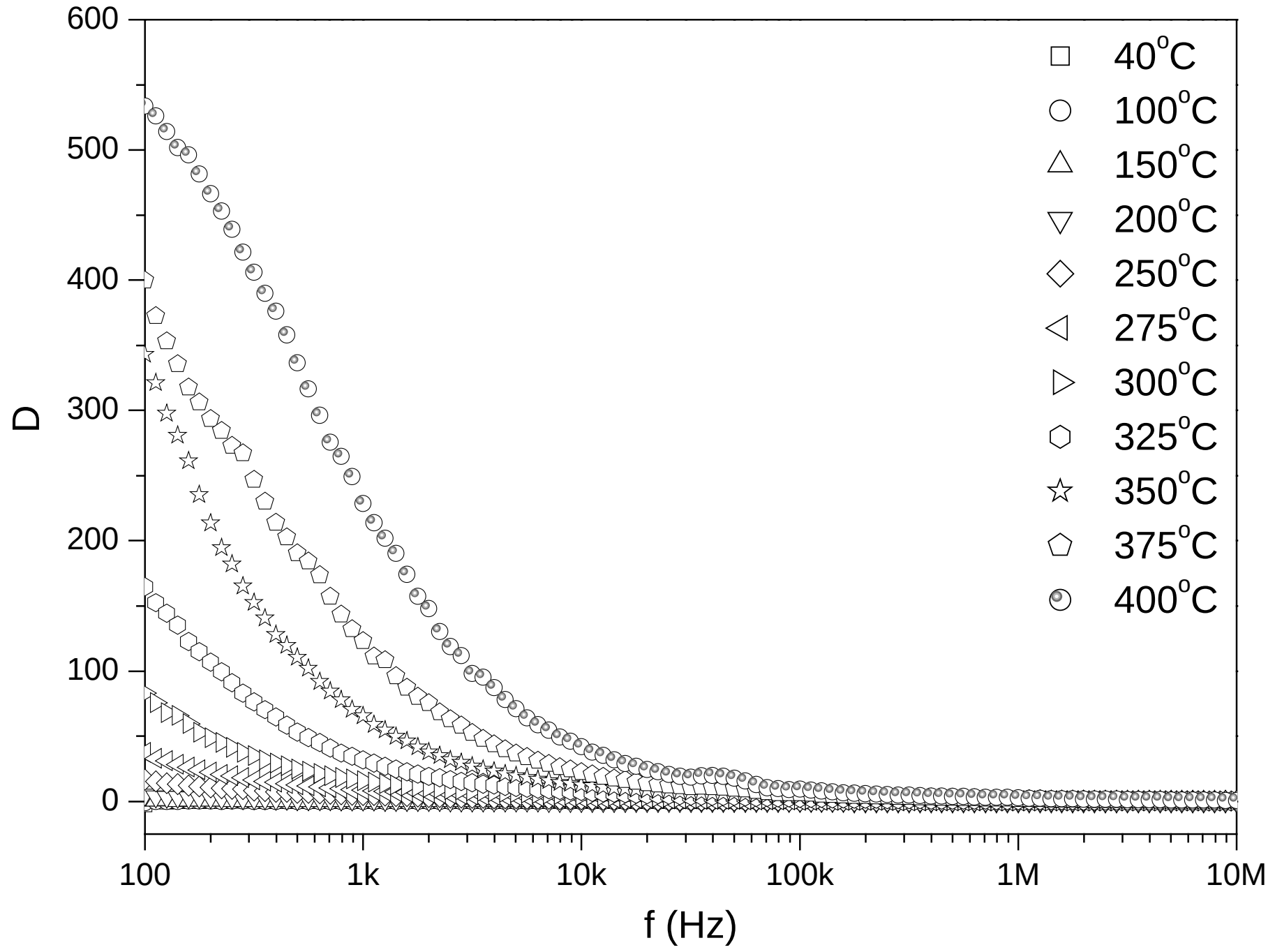

Fig. 4: Dielectric loss versus frequency plots at various temperatures.

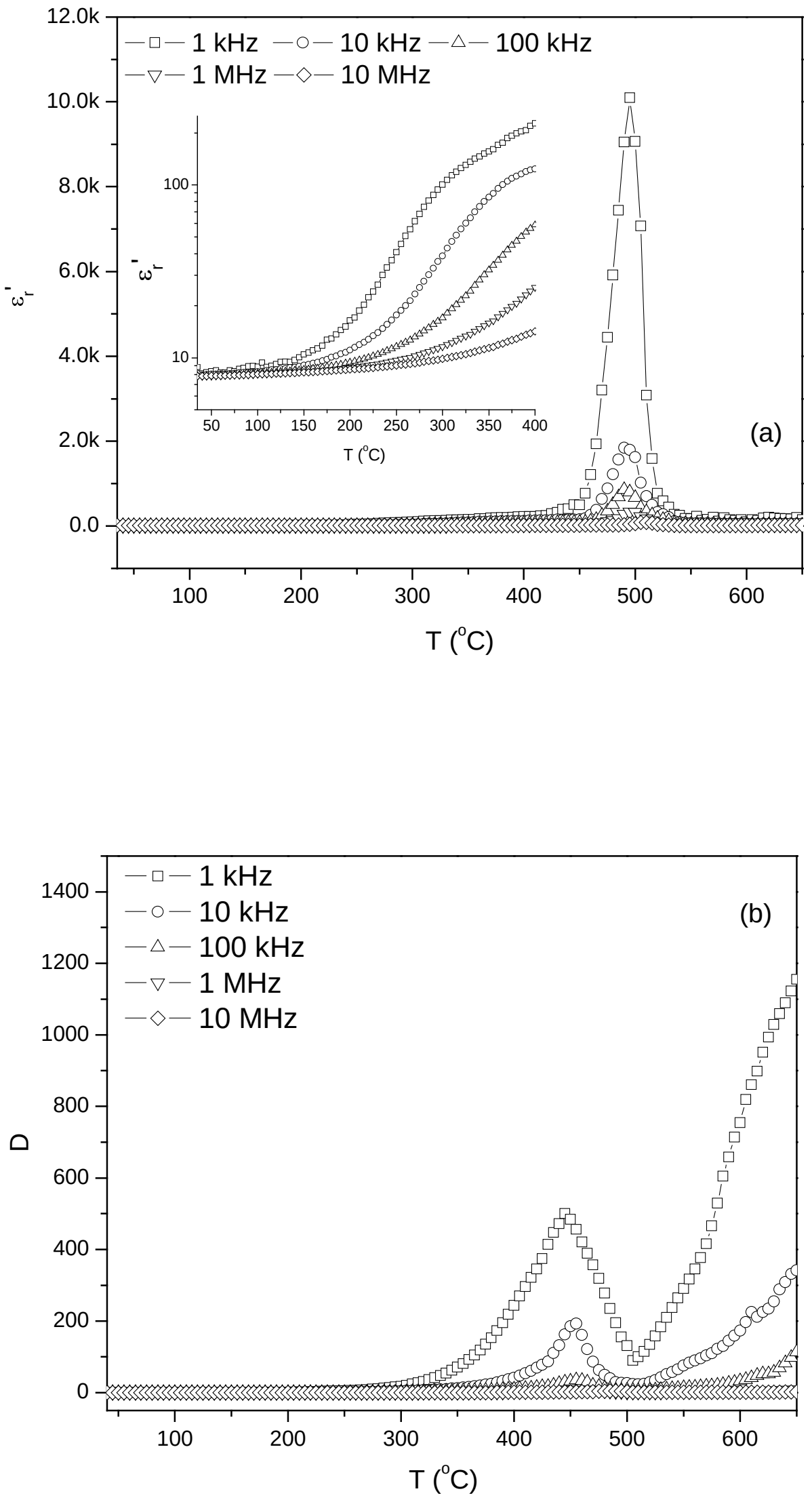

Fig. 5: Temperature dependence of the (a) dielectric constant and (b) loss at various frequencies.

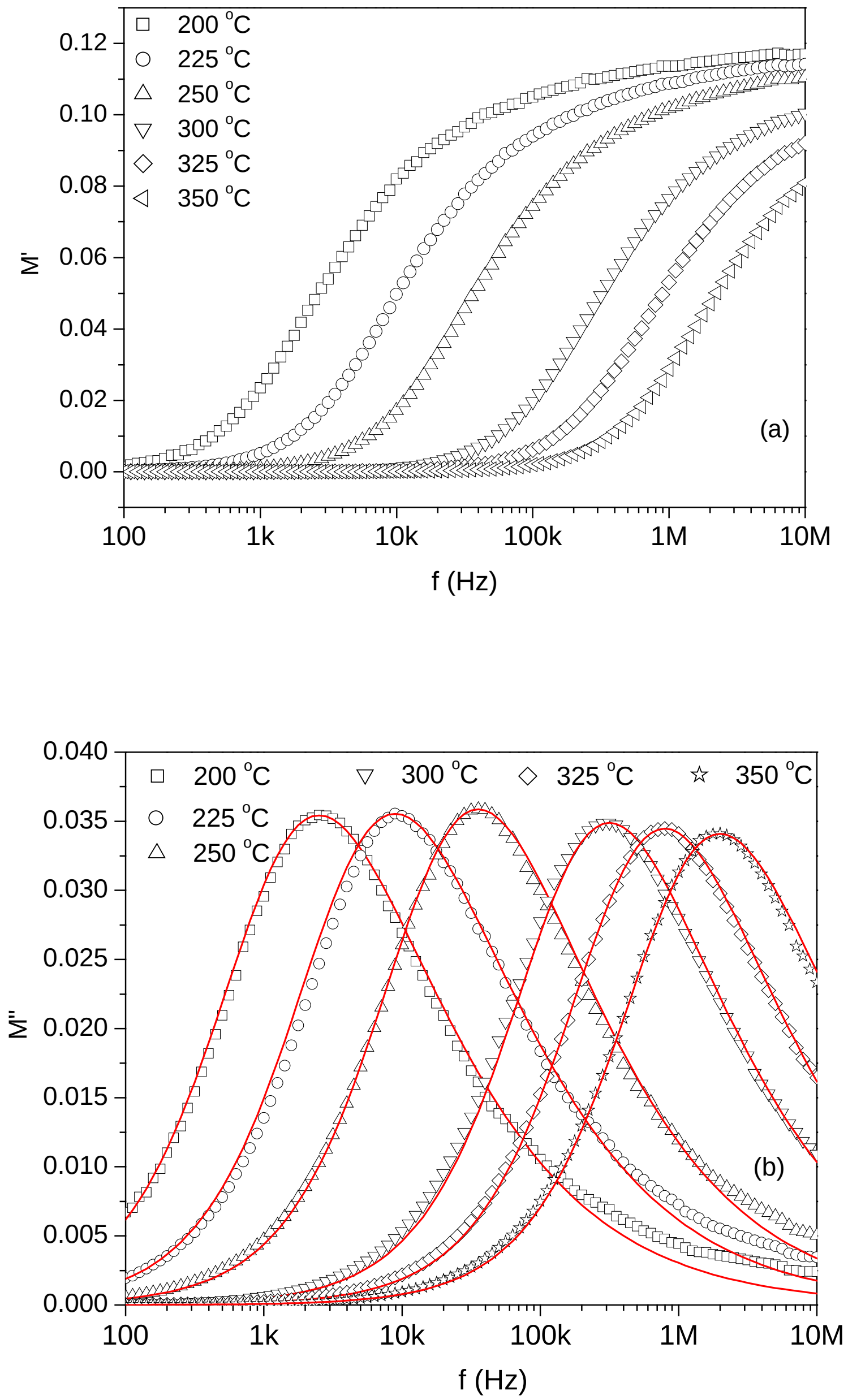

Fig. 6: (a) Real and (b) imaginary parts of the electric modulus as a function of frequency at various temperatures. The solid lines are the theoretical fits.

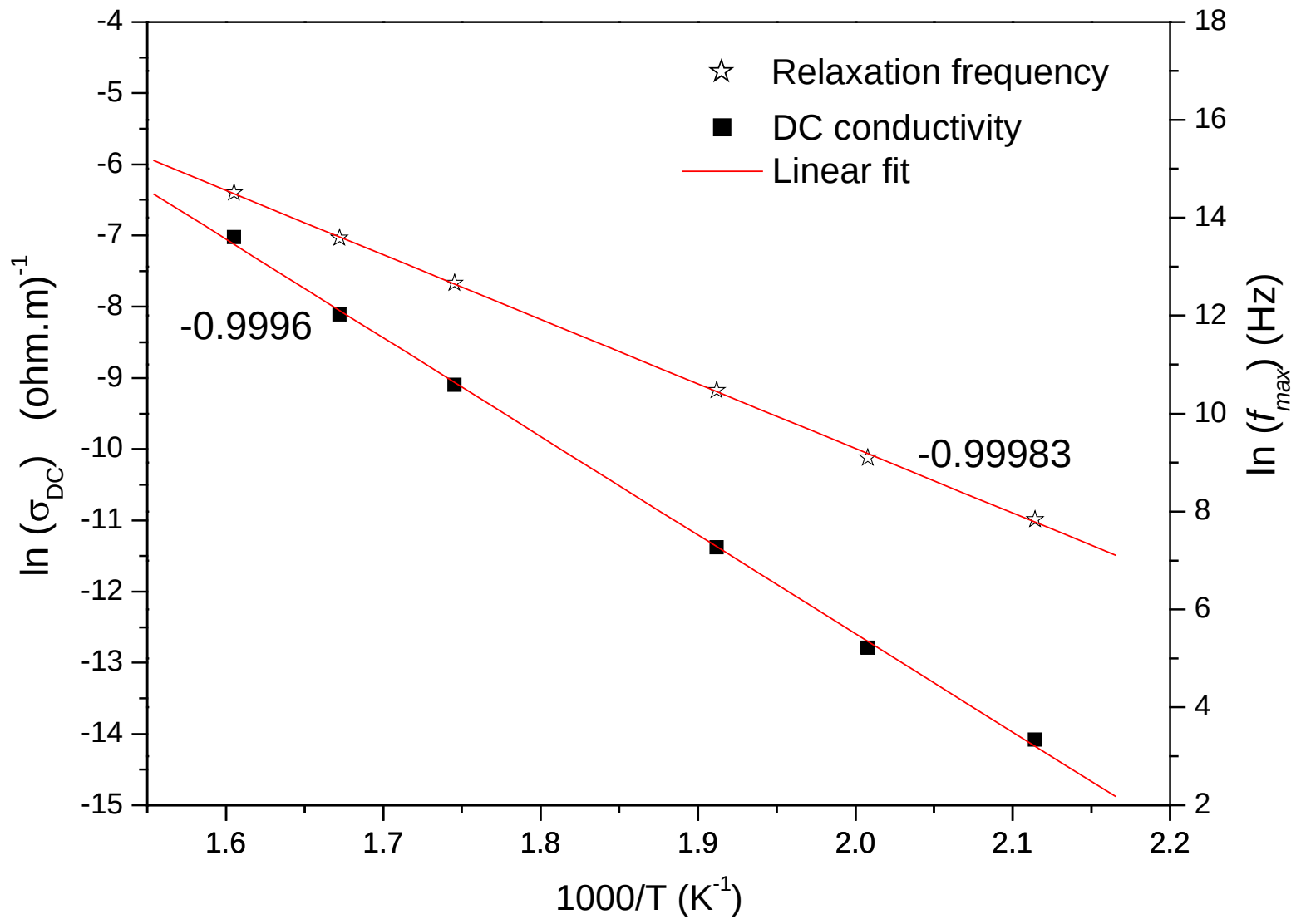

Fig. 7: Arrhenius plots of dc conductivity and electric relaxation frequency for LNBO glasses

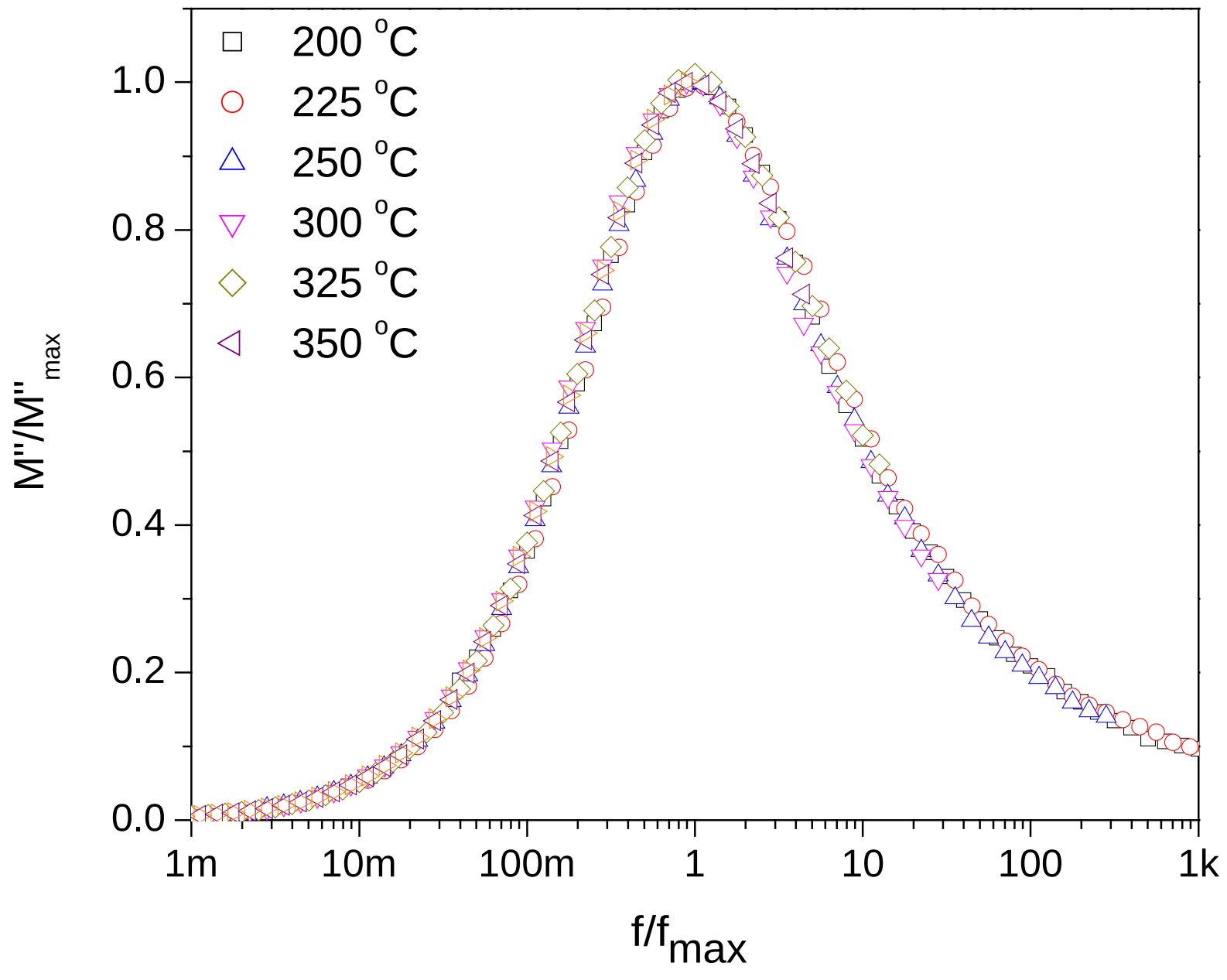

Fig. 8: Normalized spectra of electric modulus vs frequency at various temperatures

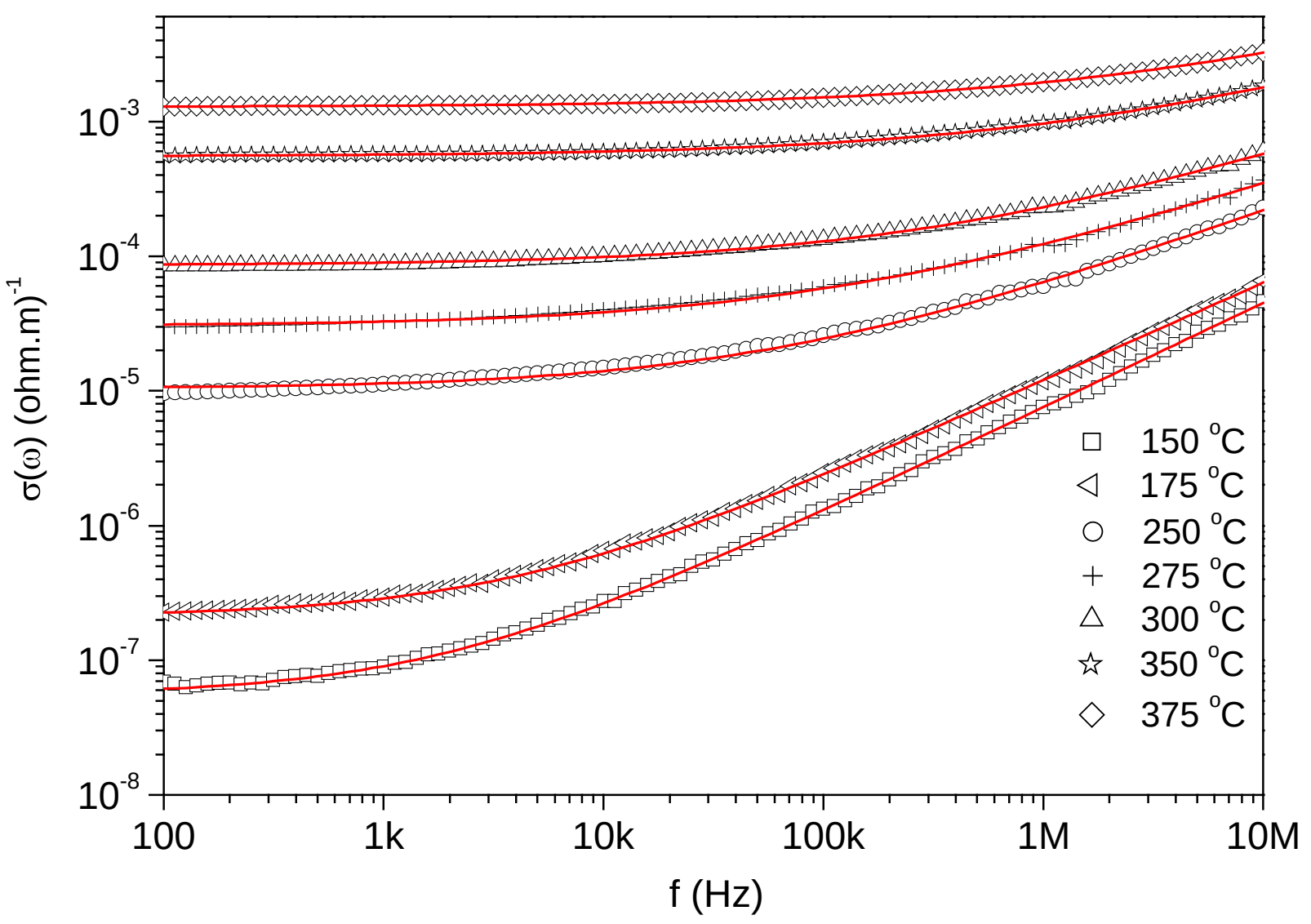

Fig. 9: Frequency dependence of ac conductivity at various temperatures and solid lines are theoretical fit

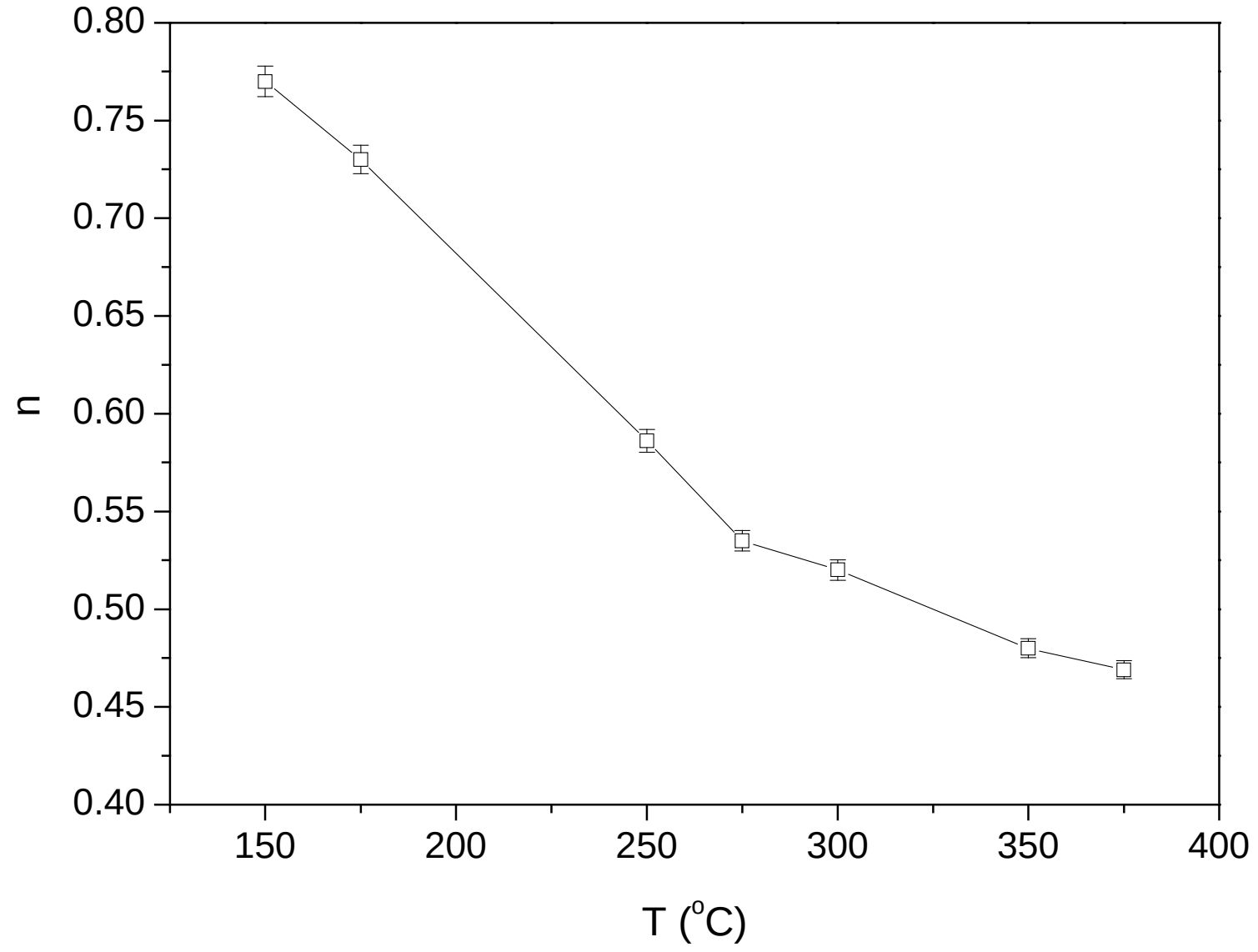

Fig. 10: Exponent ($n$) vs T plot for LNBO glasses

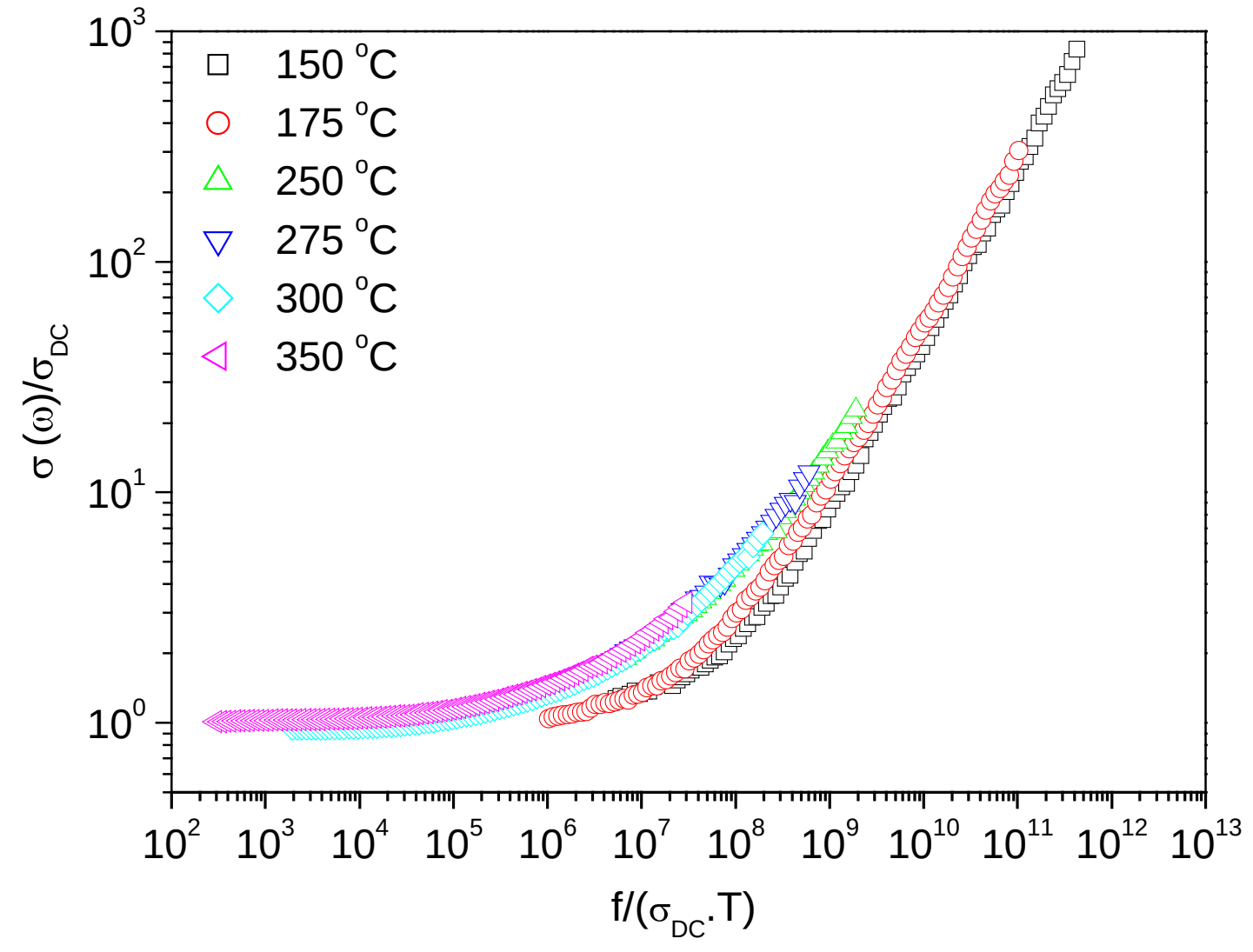

Fig. 11: Scaled conductivity spectra of LNBO glasses

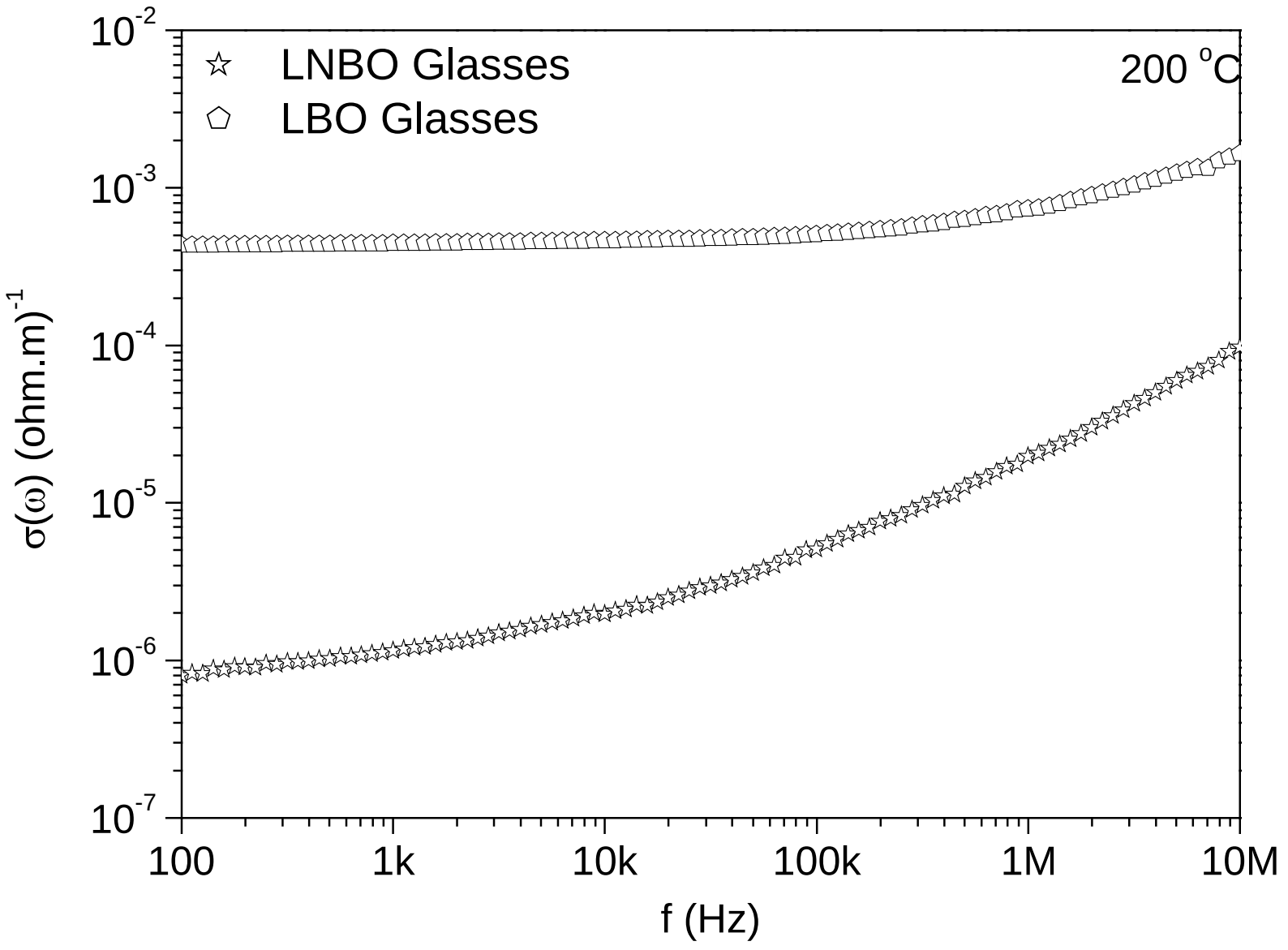

Fig. 12: Frequency dependence of conductivity of LBO and LNBO glasses